\documentclass[pre,showpacs,preprint,superscriptaddress]{revtex4}
\usepackage{amssymb,graphicx,amsbsy,amsmath}

\begin{document}

\title{Critical temperatures of the three- and four-state Potts models on the
kagome lattice}
\author{Seung Ki Baek}
\email[Corresponding author, E-mail:]{garuda@tp.umu.se}
\affiliation{Integrated Science Laboratory, Ume{\aa} University, 901 87
Ume{\aa}, Sweden}
\author{Harri M\"akel\"a}
\affiliation{Department of Physics, Ume{\aa} University, 901 87 Ume{\aa}, Sweden}
\author{Petter Minnhagen}
\affiliation{Department of Physics, Ume{\aa} University, 901 87 Ume{\aa}, Sweden}
\author{Beom Jun Kim}
\affiliation{BK21 Physics Research Division and Department of Physics,
Sungkyunkwan University, Suwon 440-746, Korea}
\affiliation{Asia Pacific Center for Theoretical Physics, Pohang 790-784, Korea}

\begin{abstract}
The value of the internal energy per spin is independent of the strip width
for a certain class of spin systems on two dimensional infinite strips. It
is verified that the Ising model on the kagome lattice belongs to this class
through an exact transfer-matrix calculation of the internal energy for the
two smallest widths. More generally, one can suggest an upper bound for
the critical coupling strength $K_c(q)$ for the $q$-state Potts model from
exact calculations of the internal energy for the two smallest strip
widths. Combining this with the corresponding calculation for the dual
lattice and using an exact duality relation enables us to conjecture the
critical coupling strengths for the three- and four-state Potts
models on the kagome lattice. The values are $K_c(q=3)=1.056~509~426~929~0$
and $K_c(q=4) = 1.149~360~587~229~2$,
and the values can, in principle, be obtained to an arbitrary precision. We
discuss the fact that these values are in the middle of earlier
approximate results and furthermore differ from earlier conjectures
for the exact values.
\end{abstract}

\pacs{64.60.De,02.70.Wz,05.70.Jk}

\maketitle

The finite-size scaling technique, known as phenomenological
renormalization, has proven to be a very reliable method for obtaining
critical properties of low-dimensional
systems~\cite{night,barber,burkhardt}.
Since its beginning, this method has been used to extract thermodynamic
quantities in the infinite-width limit from transfer-matrix calculations for
infinite strips of finite width~\cite{night}. Since a transfer-matrix
calculation gives exact results for an infinite strip with width $L$, it can
give very precise information on the system in the thermodynamic limit,
provided that the relevant thermodynamic functions have good convergences
with respect to the width $L$. For this reason, corrections to the critical
finite-size scaling have been a key issue in this phenomenological
renormalization method~\cite{luck,singh}. The present investigation 
makes use of the recent progress in computing algorithms, which makes it
possible to solve an eigenvalue problem for a transfer matrix with a
considerable size in an exact or arbitrarily precise manner. This means that
one may use symbolic algebra systems to solve a given transfer matrix in a
closed form. Alternatively, if this is not possible, one can do numerical
calculations with an arbitrary precision, which means that the numerical
precision of every calculation is free from rounding errors but limited only
by the computing memory. This makes it possible to take full advantage of
the exactness of the transfer-matrix method. In the present work we use
exact calculations of narrow infinite strips to locate the critical point
of the $q$-state Potts models~\cite{potts} on the kagome lattice with $q=3$
and $4$. The kagome lattice is one of the simplest two-dimensional (2D)
structures belonging to the Archimedean lattices and has also drawn
practical attention due to distinct structural properties~\cite{kag1,kag2}.
The case of $q=2$ on the kagome lattice was solved more than half a century
ago~\cite{kano}, but the three- and four-state Potts models have been
long-standing open questions in statistical physics, and have given rise to,
by now, classical conjectures~\cite{wu,tsallis}, as well as a number of
numerical and approximate determinations~\cite{jensen,chen,wu2010,ding}.

The zero-field $q$-state Potts model is defined by the following
Hamiltonian:
\[ H = -J \sum_{\left< ij \right>} \delta(S_i, S_j), \]
where each spin $S_k$ may take an integer value from $0$ to $(q-1)$,
$\delta$ denotes the Kronecker delta function, and the sum is over all the
nearest-neighbor pairs. We will set the interaction strength $J$ as unity
throughout this work and identify the inverse temperature $\beta$ with the
coupling strength $K \equiv \beta J$.
According to the Fortuin-Kasteleyn representation~\cite{fk}, the partition
function corresponding to this Potts Hamiltonian can be written as
\begin{equation}
Z = \sum_{\{S\}} e^{-\beta H} = \sum_{\{S\}} p^b (1-p)^{B-b} q^{N_c}
\label{eq:fk}
\end{equation}
with $p \equiv 1-e^{-K}$,
where the sum is over all the spin configurations with $N_c$ clusters made
of $b$ connected bonds out of $B$ total bonds inside the system. At the
critical point $K_c$ the partition function for the infinite system (=both
length and width infinite) has singularities in its $K$-derivatives. The
conjecture in Ref.~\cite{wu} states that the critical points can be located
by solving the following sixth-order polynomial:
\begin{equation}
v^6 + 6v^5 + 9v^4 - 2qv^3 - 12qv^2 - 6q^2v - q^3 = 0
\label{eq:wu}
\end{equation}
with $v \equiv e^K-1$. As will be described below, our estimates, based on the
two thinnest infinite strips, are very close to the values predicted by
this conjecture.

\begin{figure}
\includegraphics[width=0.15\textwidth]{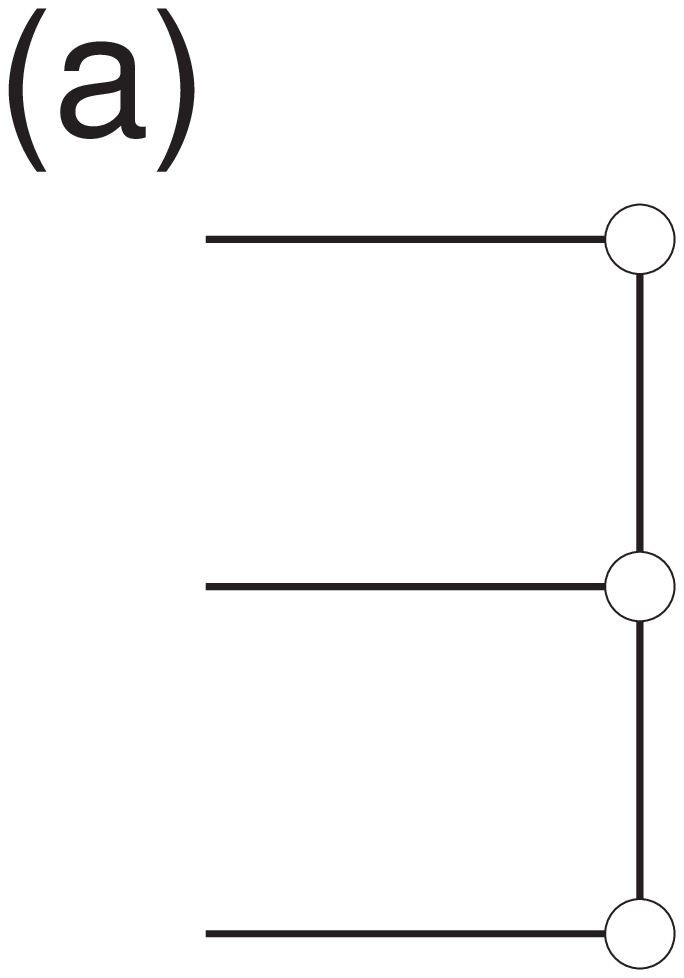}
\includegraphics[width=0.15\textwidth]{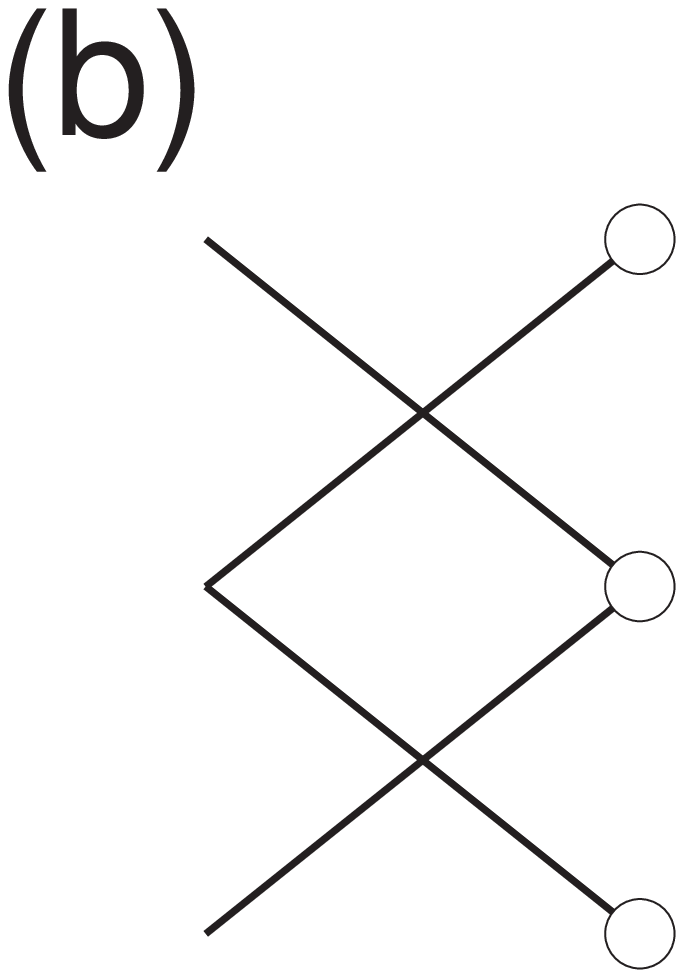}
\caption{Spin blocks to make spin strips of (a) the square-lattice type,
and (b) the double-square-lattice type.
The periodic boundary condition is imposed in the vertical direction for
all the cases, so the vertical lengths are regarded as $L=3$.
Note that the periodic boundary condition may introduce {\em double}
connections in some pairs of spins if $L$ is small.
}
\label{fig:bl}
\end{figure}

To illustrate the transfer-matrix method~\cite{binney}, we first consider a
thin strip of spins with size $\infty \times L$ as shown in
Fig.~\ref{fig:bl}(a). Once the transfer matrix is obtained, the free energy
per spin $f_L$ is given in terms of the largest eigenvalue $\lambda_L^{(0)}$
of the matrix as
\begin{equation}
-\beta f_L = L^{-1} \log \lambda_L^{(0)},
\label{eq:f}
\end{equation}
and the internal energy per spin is therefore given as
\begin{equation}
u_L = \frac{\partial}{\partial \beta} (\beta f_L) = -
\frac{1}{L \lambda_L^{(0)}} \frac{\partial \lambda_L^{(0)}}{\partial \beta}.
\label{eq:u}
\end{equation}
Furthermore, given the second largest eigenvalue $\lambda_L^{(1)}$,
the inverse correlation length is obtained as
$\xi_L^{-1} = \log \frac{\lambda_L^{(0)}}{|\lambda_L^{(1)}|}$.
In the context of the arbitrary-precision arithmetic, the differentiation in
Eq.~(\ref{eq:u}) may need some care. The derivatives of the eigenvalues
can be calculated by using the equation $\Lambda' = Y^{\ast} T' X$,
where $T' = \partial T / \partial \beta$ is the first-order derivative of
the matrix $T$ and $\Lambda$ is a diagonal matrix with the eigenvalues of
$T$~\cite{nico}. The matrices $X$ and $Y$ represent the right and left
eigenvectors, respectively, which are constructed in such a way that
$Y^{\ast}X = I$ is the identity matrix. Here, the asterisk $\ast$ means the
complex conjugate transpose.
Suppose that the spins on the strips are described by the $q$-state Potts
model with $q=2$, which is equivalent to the Ising model with the
temperature divided by $2$. It has been shown in Ref.~\cite{night} that the
correlation length is very well approximated by $\xi_L \propto L$ near
the critical coupling strength $K_c = \beta_c J=
\log(1+\sqrt{2})$~\cite{kramers,onsager}. This means that close enough to
the critical coupling strength one may use the width $L$ as a substitute for
the correlation length and describe the system in terms of this length
scale. The proportionality coefficient between $\xi_L$ and $L$ in the limit
$L \rightarrow \infty$ is related to the correlation-decay exponent
$\eta$ by conformal invariance~\cite{cardy}. In fact, it has been
furthermore found that $\left.\frac{\partial}{\partial \beta}
~\xi_L^{-1}\right|_{\beta=\beta_c} = - \xi_L^{-2} (\partial \xi_L / \partial
\beta) = const.$ for every finite $L$~\cite{yuri00}. Assuming that $\xi_L
\sim (\beta-\beta_c)^{-\nu}$ near $\beta_c$ in the limit of $L \rightarrow
\infty$, this yields the exact correlation-length exponent
$\nu=1$~\cite{yuri00}. Another interesting fact, crucial for the present
investigation, is that the internal energy per spin [Eq.~(\ref{eq:u})] has
at $\beta=\beta_c$ the same value for all the strips \emph{irrespective} of
their widths $L$~\cite{wosiek}.
This fact opens up a simple and practical way of locating the critical point
of the 2D Ising model by calculating the internal energy for the two
thinnest strips and then finding the coupling strength for which they have
the same internal energy. For the square-lattice strip in the diagonal
direction shown in Fig.~\ref{fig:bl}(b), for example, equating the internal
energy per spin for $L=2$ to that of $L=3$, we get
\begin{equation*}
\frac{e^{2K} (e^{2K} - 1) (e^{4K} - 6e^{2K} +
1)}{(e^{2K}+1) F(K) + (e^{2K}+1)^2 \sqrt{F(K)}} = 0,
\end{equation*}
where $F(K) \equiv e^{8K} - 8e^{6K} + 30e^{4K} - 8e^{2K} + 1$.
It is straightforward to see that the nonnegative solutions of the equation
are $e^K = 0, 1, \infty$, and $\sqrt{2} \pm 1$. Only the latter two are
nontrivial and give us the exact $K_c$ for the 2D ferromagnetic and
antiferromagnetic models, respectively. The largest eigenvalues for
$L=4$ and $5$ are also available in closed forms and lead to the
same conclusion. By using the geometry shown in Fig.~\ref{fig:bl}(a),
the exact results $e^K = 1 + \sqrt{q}$ have been obtained for $q
\le 5$~\cite{wosiek}, but there are also cases where the method does not
apply~\cite{wosiek,souza}. The invariance of the internal energy with
respect to strip width has therefore been conjectured to be due to certain
symmetries in the model~\cite{souza,yuri02}.

\begin{figure}
\includegraphics[width=0.15\textwidth]{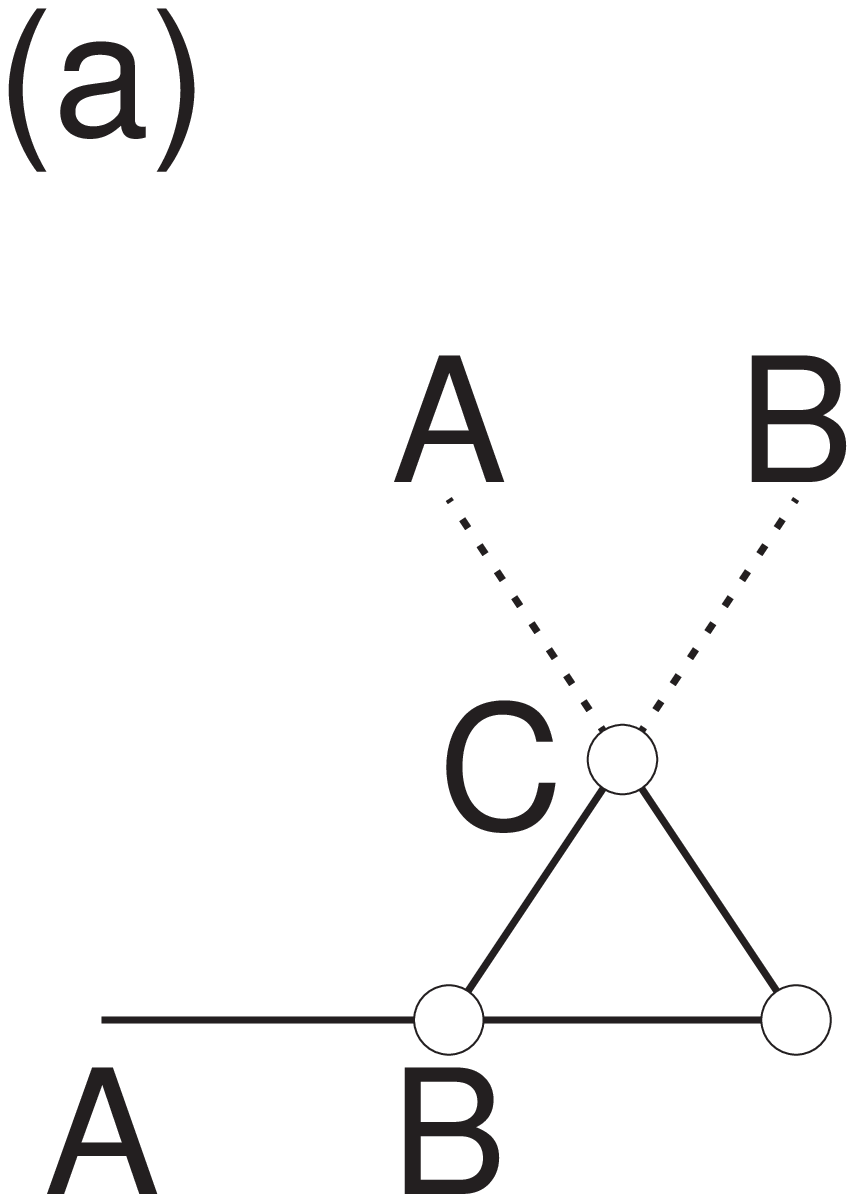}
\includegraphics[width=0.15\textwidth]{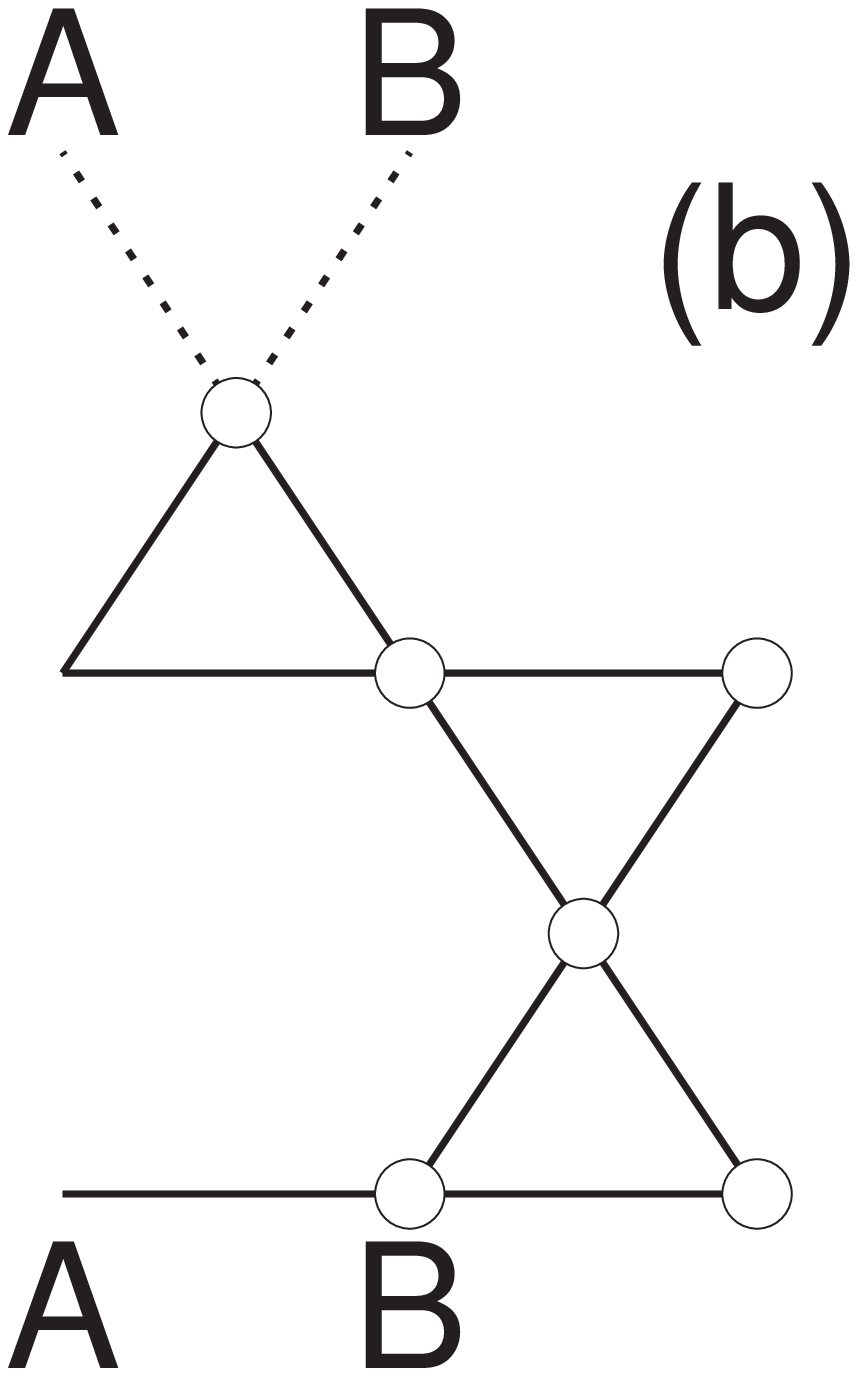}
\includegraphics[width=0.15\textwidth]{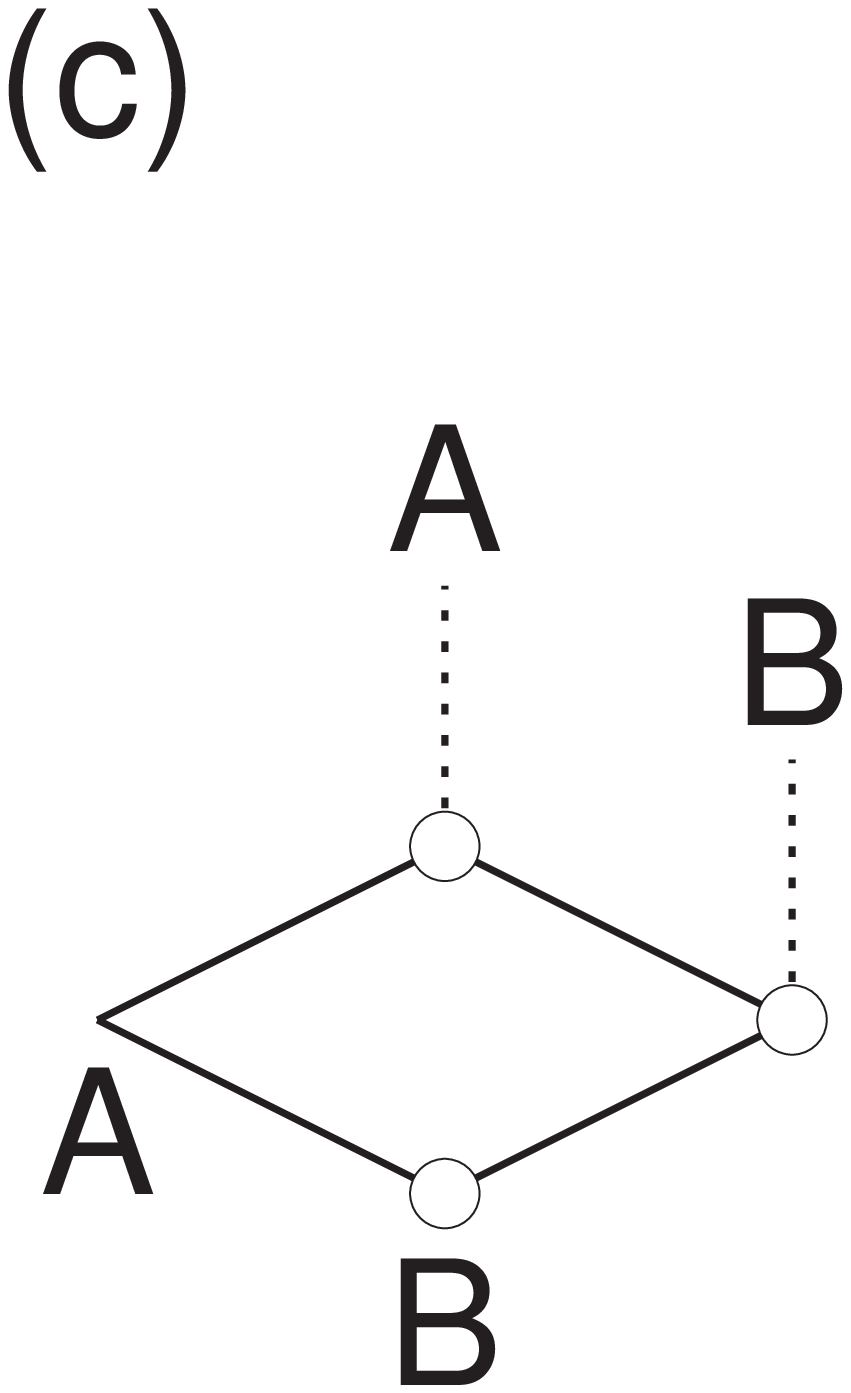}
\includegraphics[width=0.15\textwidth]{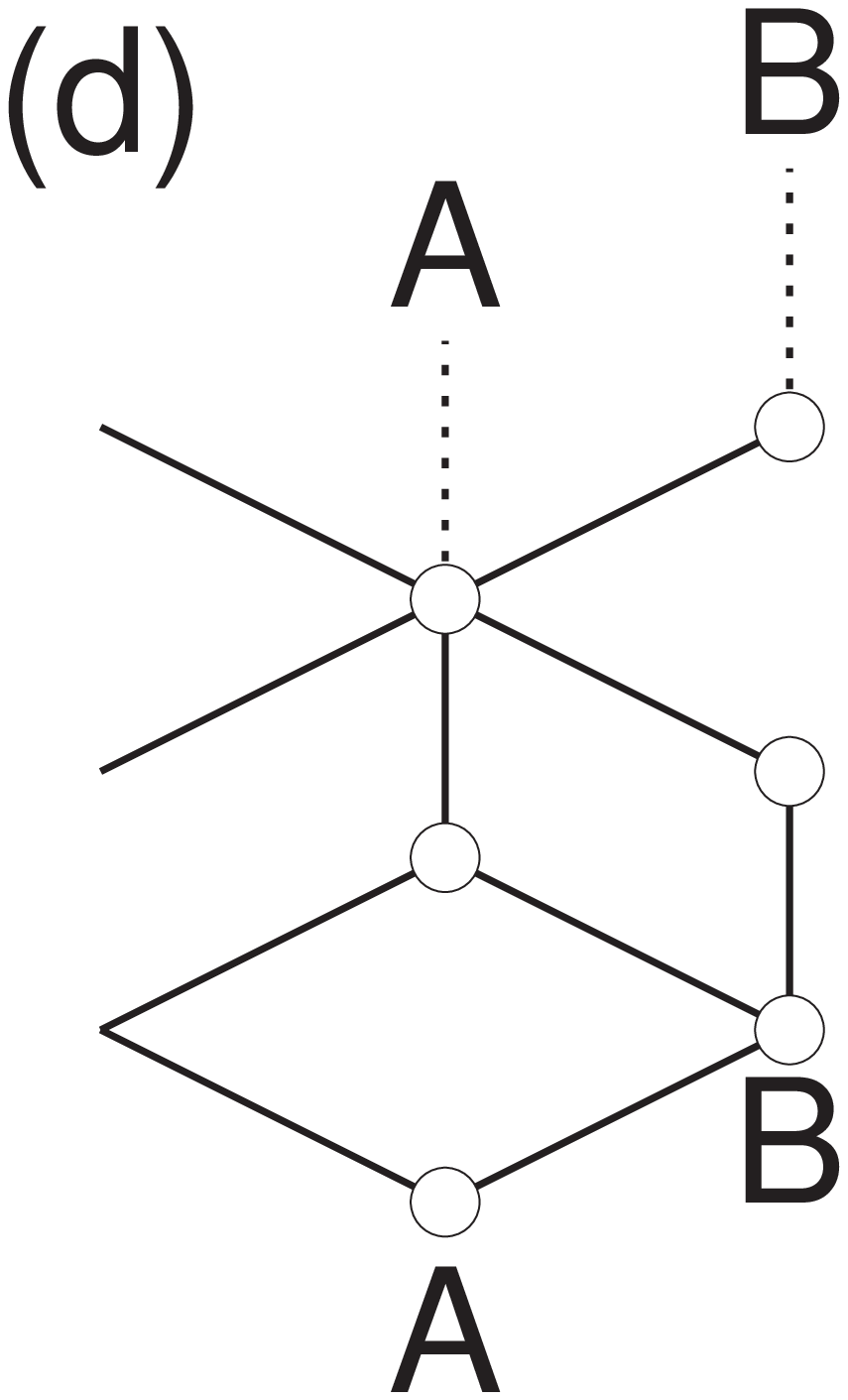}
\caption{Spin blocks to make spin strips of the kagome type with (a) $L=1$
and (b) $L=2$, and those of the dice type with (c) $L=1$ and (d) $L=2$.
The dotted lines show the periodic boundary condition in the vertical
direction. Note that $B$ and $C$ are {\em doubly} connected in (a), and such
double connections are also found in (c).
}
\label{fig:kag}
\end{figure}

We apply the transfer-matrix method to the Potts models on the kagome lattice.
First, we verify that the
known exact solution for the two-state Potts model is reproduced by assuming
that the internal energy per spin is invariant also in this case. We
construct two spin blocks for generating the kagome lattice, as illustrated
in Figs.~\ref{fig:kag}(a) and \ref{fig:kag}(b).
Writing down the corresponding transfer matrices and denoting their largest
eigenvalues as $\lambda^{(0)}_a$ and $\lambda^{(0)}_b$, respectively, we
compute the internal energies per spin as
$u_a = -(3 \lambda_a^{(0)})^{-1} (\partial \lambda_a^{(0)} / \partial
\beta)$ for Fig.~\ref{fig:kag}(a) and
$u_b = -(6 \lambda_b^{(0)})^{-1} (\partial \lambda_b^{(0)} / \partial
\beta)$ for Fig.~\ref{fig:kag}(b).
Indeed, it is readily found that $u_a = u_b = -(7+2\sqrt{3})/6$ at $\beta =
\frac{1}{2} \log(3 + 2\sqrt{3})$, which is the exact critical point of this
system~\cite{kano}. This verifies that the correct critical $K_c$ can be
obtained from the two thinnest strips also for the two-state Potts model on
the kagome lattice. Or, in other words, this shows that the internal energy
per spin is invariant also for the two-state Potts model on the kagome
lattice.

\begin{figure}
\includegraphics[width=0.45\textwidth]{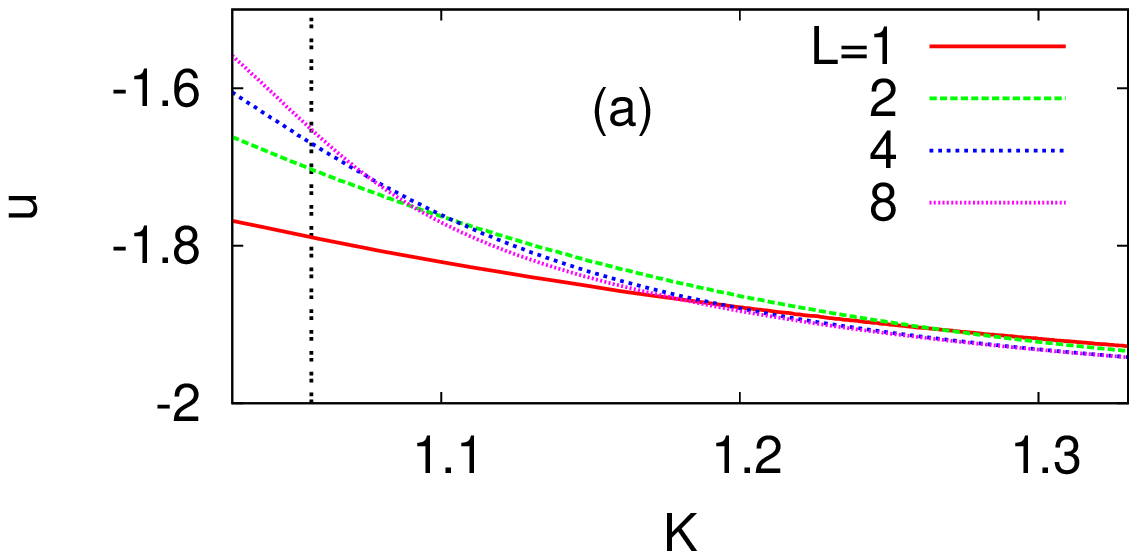}
\includegraphics[width=0.45\textwidth]{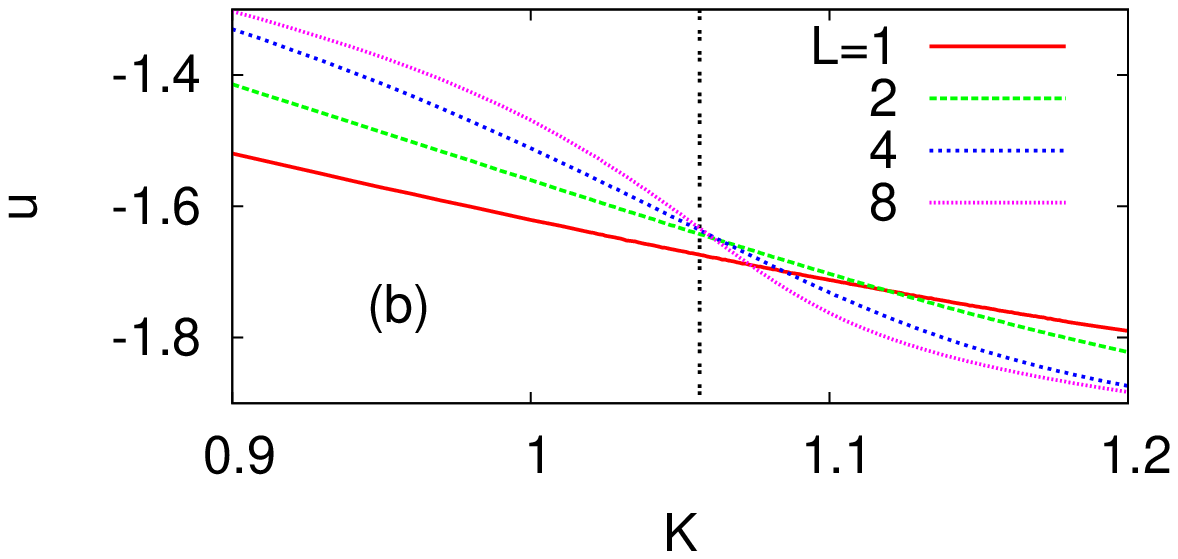}
\includegraphics[width=0.45\textwidth]{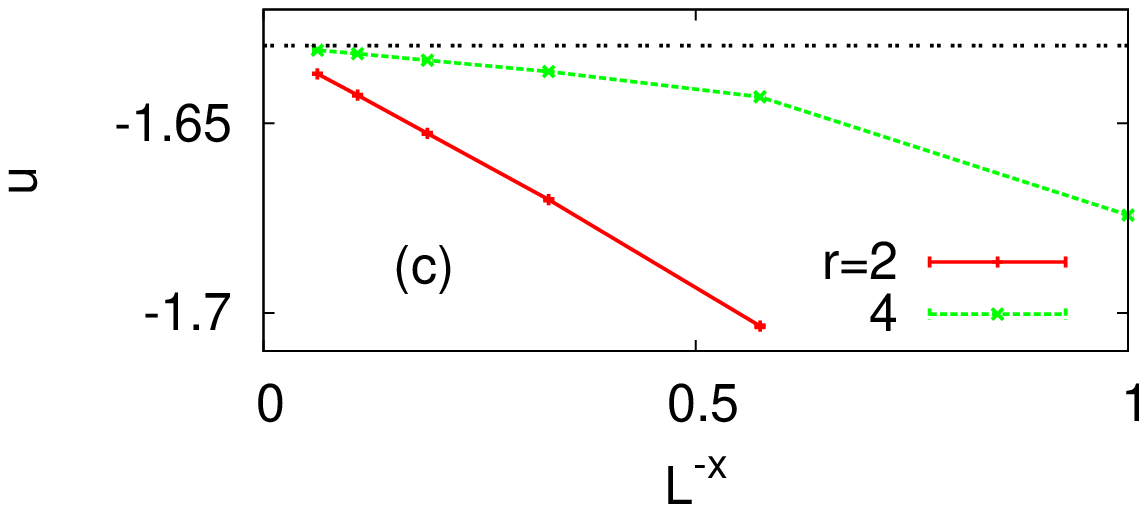}
\includegraphics[width=0.45\textwidth]{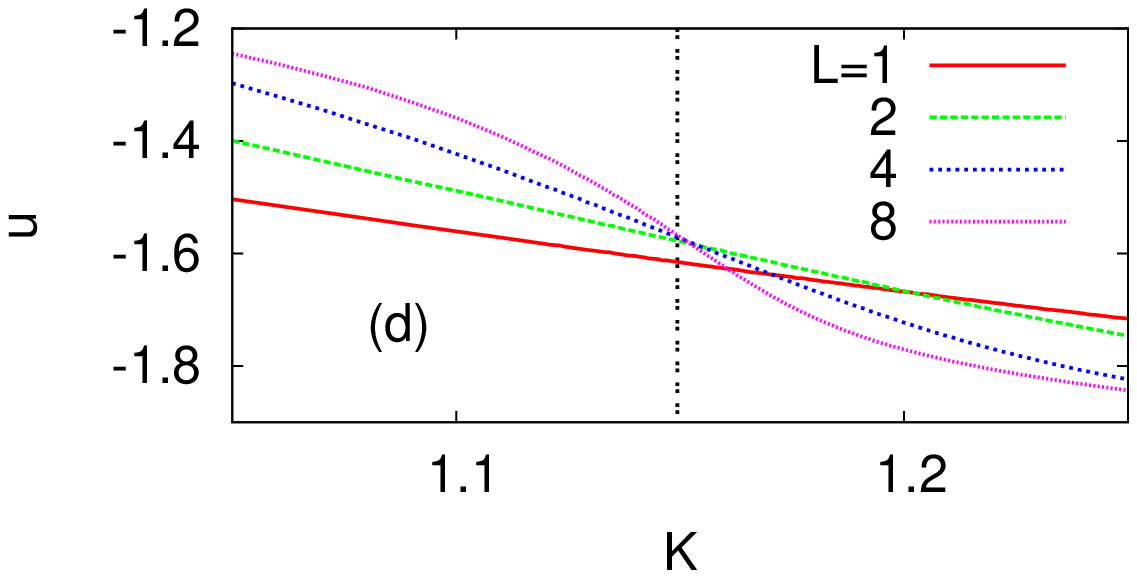}
\caption{Monte Carlo data of the internal energy per spin for $q=3$, with
aspect ratios (a) $r=2$ and (b) $r=4$, respectively. The vertical dotted
lines show the critical coupling strength obtained in this work. (c)
Sectional view at this particular point, where the horizontal dotted line
indicates the value obtained in this work, $u = -1.629~543~706~399~6$.
The scaling dimension $x$ is expressed as $(1-\alpha)/\nu$ with the
specific-heat exponent $\alpha$, and therefore $x=4/5$ for $q=3$.
(d) Qualitatively the same behavior is observed for $q=4$. Here we plot it
with $r=4$.
}
\label{fig:mc}
\end{figure}

In case of the three- and four-state Potts models on the kagome lattice,
neither the critical values $K_c$ are exactly known, nor is it \emph{a
priori} known if the internal energy is invariant. In order to generalize
the method, we study pairs of strips, one with finite length $M$ and width
$L$ and the other with $2M$ and $2L$. The two strips in such a pair have the
same aspect ratio $r=M/L$ where $M$ and $L$ are chosen such that $r$ is a
positive number. When $r$ is suitably chosen, each such pair will
have a single $K_{\rm cross}(L)$ for
which the internal energy per spin is the same. Figures~\ref{fig:mc}(a) and
\ref{fig:mc}(b) show that the crossing point $K_{\rm cross}(L)$
monotonically decreases with increasing $L$. In the limit of
$L\rightarrow\infty$, the crossing point approaches the true critical $K_c$,
that is, $[K_{\rm cross}(L) - K_c] \rightarrow 0^+$.
This implies that $K_{\rm cross}(L)$
will give an upper bound for $K_c$ for each of the fixed aspect ratios.
As long as the aspect ratio does not change any essential physics but only the
convergence rate toward the bulk criticality [Fig.~\ref{fig:mc}(c)],
we can suggest that the crossing point $K_{\rm
cross}(L=1)$ for $r=\infty$ will either give an upper bound or alternatively
the exact results: that is, for the case of $q=2$ it gives the exact result,
whereas it gives at least an upper bound for $q=3$ and $q=4$.
An argument can be given in the following way:
the crossing point would fail to be an upper bound
if crossing could be found on
both sides of the true critical point $K_c$. This actually means that the
internal energy per spin $u_L$ would not be a monotonic function of $L$ at
$K=K_c$. We note that
the classical Potts model on the $L \times M$ strip can be mapped to the
one-dimensional quantum Potts model of size $L$
by putting the strip length $M$ in the
imaginary-time direction~\cite{senthil}. One
can describe the finite-size scaling around the critical point
as $u_L - u_{\infty} = L^{-x} a \left[ (K - K_c) L^{1/\nu}, M L^{-z}
\right]$ to the leading order with a two-parameter function
$a$~\cite{young}, the dynamic critical exponent $z=1$~\cite{senthil}, and
the scaling dimension $x$ of the energy-density operator~\cite{weigel}.
At $K=K_c$, the scaling function reduces to $a(ML^{-1}) = a(r)$, so we find
that $u_L \sim u_{\infty} + a(r) L^{-x}$. The geometric factor $r$ is
absorbed by the coefficient $a$ which determines the convergence rate.
It would be plausible to say that $a(r)$ is continuous and nonvanishing for
any finite $r$ and hence cannot change the sign. The theory of the
finite-size scaling therefore tells us that $u_L$ is a monotonic function of
$L$ so that the crossing point for a fixed aspect ratio will exist only on
one side, which is $K \ge K_c$ in this case.

In the present investigation we are for practical reasons restricted to
$q\le 4$ and $L \le 2$ since the transfer-matrix size increases as $q^{3L}
\times q^{3L}$. It is straightforward to write the transfer matrices for
$q=3$ and $4$ and solve the eigenvalue problem. By equating $u_a$ to $u_b$
as above, we obtain the two values $K_{\rm cross}(L=1)$, which are
$1.056~509~426~929~0$ and  $1.149~360~587~229~2$ for $q=3$ and $q=4$,
respectively. These values are shown in Table~\ref{table:val} together with
other existing estimates. Note that the arbitrary-precision arithmetic can
make our values as precise as we want, in principle. As seen in
Table~\ref{table:val}, our values are somewhere in the middle of the earlier
existing estimates and conjectures, suggesting that they may be the exact
values.
In order to further examine the obtained values, we make
use of the fact that there exists an exact relation between the critical
coupling strengths of the kagome lattice and its dual [called a dice
lattice, compare Figs.~\ref{fig:kag}(c) and \ref{fig:kag}(d)]~\cite{wu,wu77},
\begin{equation}
(e^{K_c}-1)(e^{\tilde{K}_c}-1)=q,
\label{eq:d}
\end{equation}
where $\tilde{K}_c$ means the critical coupling strength of the dice lattice.
This means that the upper bound $\tilde{K}_{\rm cross}(L=1)$ obtained for the
dual lattice can be turned into a \emph{lower} bound for $K_c$ of the 
kagome lattice. Repeating the calculation for $\tilde{K}_{\rm cross}(L=1)$
with the two thinnest strips, given in Figs.~\ref{fig:kag}(c) and
\ref{fig:kag}(d), gives $\tilde{K}_{\rm cross}(L=1)=0.955~080~568~397~4$ as
an upper bound for $\tilde{K}_c$, which through Eq.~(\ref{eq:d}) gives the
\emph{lower} bound $K_c=1.056~509~426~929~0$. This is, to all the 14
decimal places, identical to the \emph{upper} bound obtained directly for
the kagome lattice. The most reasonable conclusion is that the calculation
gives the exact value and that, just as for the two-state Potts model on the
kagome lattice, the internal energy is independent of the strip width at
the critical temperature and that, furthermore, the same is true for the
$q=4$ case [Fig.~\ref{fig:mc}(d)]. A complete analytic argument is called
for, and a simple way to test this conjecture would be to solve the transfer
matrix with $L=3$.

As seen from Table~\ref{table:val}, the situation for the three-state Potts
model is as follows: both the two earlier conjectured exact values can be
ruled out, although the conjecture by Wu in Ref.~\cite{wu} is very close to
the value in this work. Our conjectured exact
value is somewhat surprisingly outside the bounds
of the value estimated from series expansion in Ref.~\cite{jensen} and the
subnet estimate in Ref.~\cite{ding}. It agrees well with and is inside the
bounds of the Monte Carlo estimate in Ref.~\cite{chen}. For the four-state
Potts model, the situation is somewhat different: again, the earlier conjectured
exact values can be ruled out. However, our conjectured exact value is
inside the bounds of all the other estimates.
 
\begin{table}
\caption{Critical thresholds of the $q$-state Potts models on the
kagome lattice in terms of $p = 1-e^{-K}$.}
\begin{tabular*}{\hsize}{@{\extracolsep{\fill}}ccccc}\hline\hline
Reference & $q=1$ & $q=2$ & $q=3$ & $q=4$ \\\hline
Exact~\cite{kano} & & $1-1/\sqrt{3+2\sqrt{3}}$ & & \\
Conjecture~\cite{wu} & $0.524~429~71$ & $1-1/\sqrt{3+2\sqrt{3}}$ & $0.652~327~40$ & $0.683~127~34$ \\
Conjecture~\cite{tsallis} & $0.522~372~07$ & $1-1/\sqrt{3+2\sqrt{3}}$ & $0.653~932~82$ & $0.685~967~83$ \\
Series~\cite{jensen} & & & $0.652~350(5)$ & $0.683~15(5)$ \\
Monte Carlo~\cite{ziff} & $0.524~405~3(3)$ & & & \\
Monte Carlo~\cite{chen} & & $0.606~62(8)$ & $0.652~32(7)$ & $0.683~17(2)$ \\
Subnet~\cite{ding} & $0.524~404~978(5)$ & $0.606~680~106~83(15)$ & $0.652~350~2(4)$ & $0.683~163(5)$ \\
This work & & $1-1/\sqrt{3+2\sqrt{3}}$ & $0.652~332~747~264~01$ &
$0.683~160~704~284~84$ \\\hline\hline
\end{tabular*}
\label{table:val}
\end{table}

The conjecture by Wu in Ref.~\cite{wu} gives the critical coupling strengths
as solutions of the sixth-order polynomial given in Eq.~(\ref{eq:wu}).
It is important to note that we have also given our values as solutions of
certain polynomial equations since we are dealing with transfer matrices.
Although we have not factorized the full polynomials yet,
one may ask if such sixth-order polynomials as Wu has derived
can be eventually factored out.
To answer this, we follow Ref.~\cite{ziff} and try to determine
polynomials in the variable $v = e^K-1$ which have roots at the exact
critical values. Even if we work with numeric values, instead of symbolic
manipulations, this method makes it possible to find such a polynomial. For
example, in case of $q=2$, one recovers the compact analytic expression $v
=\sqrt{3+2\sqrt{3}}-1$ by solving the obtained polynomial equation. Based on
the conjecture by Wu, we try to find the value for the $q=3$ case as
the solution of the sixth-order polynomial $\sum_{i=0}^6 c_n v_n^i=0$ with
integer-valued coefficients. We furthermore assume that $c_6=1$, and let
$c_5$ and $c_4$ vary from $-25$ to $25$, while the other four coefficients
may take values from $-10^2$ to $10^2$. Substituting our value $v =
1.876~313~463~895$ for $q=3$, the best polynomial is found to be
\[ v^6 - 6v^5 + 22v^4 - 79v^3 + 99v^2 + 28v - 56= 0 ,\]
yielding a solution $v^{\rm poly} = 1.876~313~463~898$. Even if
the discrepancy between our conjectured exact value and the solution of the
polynomial is tiny, it is still significant, which means that there is no
such polynomial within the range of coefficients tested. This might suggest
that the solution cannot be obtained from a simple sixth-order polynomial as
was assumed in the conjecture by Wu.

We conclude this work with a brief sideline: the Fortuin-Kasteleyn
representation [Eq.~(\ref{eq:fk})] for $q=1$ recovers the bond-percolation
problem on the kagome lattice. The method in the present paper cannot be
directly used in this case, since the internal energy becomes a
constant independent of the coupling strength. Without the knowledge of the
exact $q$-dependence of $K_c$, one can only interpolate it from the other
estimates. For the case of the square lattice, which has coordination number
$4$, as does the kagome lattice, we have a general expression of the
critical point as $v^{\rm sq} = \sqrt{q}$~\cite{potts}. We assume that
$v(q)$ of the kagome lattice can be expanded in series of this variable:
$v(q) = a (\sqrt{q})^3 + b (\sqrt{q})^2 + c\sqrt{q}$, where we further note
that $v(0)=0$ is an exact limit. Finding the three parameters $a$, $b$, and
$c$ by substituting the conjectured values for $v(2)$, $v(3)$ and $v(4)$, we can
interpolate the value at $q=1$ and obtain $p \approx 0.52433$. Compared to
the numerical estimate shown in Table~\ref{table:val}, the fractional error
amounts to be about 140 parts per million.

In summary, we have conjectured the exact values of critical temperatures for the
three-state and four-state Potts models on the kagome lattice by using
exact transfer-matrix calculations on thin infinite strips. This suggests
that the internal energy can provide a sharper condition for criticality
than lattice symmetries considered in the earlier conjecture by Wu.
It has also been noted that, for the three-state Potts model on the kagome
lattice, the series expansion in Ref.~\cite{jensen} does not contain
our result within its bounds. The method devised to obtain the results is
based on exact solutions of the two thinnest infinite strips. These
solutions have been obtained by taking full advantage of computational
symbolic algebra systems. Since the method itself appears to be quite
general, it may possibly be used to solve other problems.

\acknowledgments
S.K.B. and P.M. acknowledge the support from the Swedish Research Council
with Grant No. 621-2008-4449.
H.M. thanks E. Lundh for access to computational resources.
B.J.K.  was supported by Basic Science Research Program through the
National Research Foundation of Korea (NRF) funded by the Ministry of
Education, Science and Technology (2010-0008758).
This research was conducted using the resources of High Performance
Computing Center North (HPC2N).


\end{document}